\documentclass{aa}  

\usepackage{natbib}
\usepackage{graphicx}
\usepackage{txfonts}
\usepackage{hyperref}

\usepackage{bm}

\newcommand{\planck}{{\it Planck}}

\newcommand{\xmm}{{\it XMM-Newton}}

\newcommand{\beq}{\begin{equation}}
\newcommand{\eeq}{\end{equation}}
\newcommand{\beqa}{\begin{eqnarray}}
\newcommand{\eeqa}{\end{eqnarray}}

\newcommand{\msunh}{$h^{-1} {\rm M}_{\odot}$}

\newcommand{\msun}{${\rm M}_{\odot}$}

\newcommand{\eg}{e.g.,\xspace}

\def\msun{{\rm M}_{\odot}}
\def\der{{\rm d}}

\begin{document} 

  \title{Testing decaying dark matter models as a solution to the $S_8$ tension \\ with the thermal Sunyaev-Zel'dovich effect
  }
  \titlerunning{Solving the $S_8$ tension with the SZ power spectrum}
  \author{Hideki Tanimura\inst{1,2} \and Marian Douspis\inst{1} \and Nabila Aghanim\inst{1} \and Joseph Kuruvilla\inst{1} }

  \institute{
    Universit\'{e} Paris-Saclay, CNRS, Institut d'Astrophysique Spatiale, B\^atiment 121, 91405 Orsay, France \and
    Kavli IPMU (WPI), UTIAS, The University of Tokyo, Kashiwa, Chiba 277-8583, Japan \\
    \email{hideki.tanimura@ipmu.jp} }
  \date{}

  \abstract 
    {Considering possible solutions to the $S_8$ tension between the \planck\ cosmic microwave background (CMB) measurement and low-redshift probes, we extended the standard $\Lambda$CDM cosmological model by including decay of dark matter (DDM). We first tested the DDM model in which dark matter decays into a form of noninteracting dark radiation. Under this DDM model, we investigated the impacts of DDM on the Sunyaev Zel'dovich (SZ) effect by varying the decay lifetime, $\Gamma^{-1}$, including the background evolution in cosmology and the nonlinear prescription in the halo mass function. We performed a cosmological analysis under the assumption of this extended cosmological model by combining the latest high-redshift \planck\ CMB measurement and low-redshift measurements of the SZ power spectrum as well as the baryonic acoustic oscillations (BAO) and luminosity distances to type Ia supernovae  (SNIa). Our result shows a preference for $\Gamma^{-1} \sim 220$ Gyr with a lower bound on the decay lifetime of $\sim$ 38 Gyr at 95\% confidence level. Additionally, we tested the other DDM model in which dark matter decays into warm dark matter and dark radiation. This model supports $\Gamma^{-1} \sim 137$ Gyr to resolve the $S_8$ tension with a lower bound on the decay lifetime of $\sim$ 24 Gyr at 95\% confidence level. Comparing these two models, we find that the second leads to slightly better reconciliation of the $S_8$ tension.}
    \keywords{galaxies: clusters: general - intracluster medium - Cosmology: large-scale structure of Universe - dark matter - cosmic background radiation}

\maketitle


\section{Introduction}
\label{sec:intro}

The current standard cosmological model, called the $\Lambda$ cold dark matter ($\Lambda$CDM) model, has been supported by many observational results at different epochs, such as cosmic microwave background (CMB), big bang nucleosynthesis (BBN), and luminosity distances to type Ia supernovae (SNIa). However,  as measurement precision has increased over the past few years, the validity of the $\Lambda$CDM model is under reexamination, particularly due to a discrepancy in the $S_8 (\equiv \sigma_8 (\Omega_{m} / 0.3)^{0.5}$ parameter, which is the amplitude of matter density fluctuations, $\sigma_8$, scaled by the square root of the matter density, $\Omega_m$. For example, \planck\ measured $S_8 = 0.830 \pm 0.013$ with the CMB anisotropies ($z\sim1100$) \citep{Planck2020VI}, but low-redshift ($z\sim0-1$) cosmological probes, such as the abundance of galaxy clusters by \planck\  \citep{Planck2014XX, Planck2016XXIV} and gravitational lensing by the Kilo Degree Survey (KiDS; \citealt{Heymans2021}) and Dark Energy Survey (DES; \citealt{DES2022}), show a preference for a lower value of $S_8 \sim 0.76-0.78$, representing a tension of up to $\sim2-3\sigma$ with the CMB constraints, referred to as the $S_8$ tension. This tension indicates that the growth rate of the cosmic structure is less than expected from the CMB measurements and may demand modifications to the standard model. 

Many attempts have been made to solve the $S_8$ tension by breaking assumptions in the standard cosmological model, mainly on the nature of neutrinos (\eg \citealt{Salvati2018, Troster2021}), dark energy (\eg \citealt{Bolliet2018, Salvati2018, Lambiase2019, Valentino2020, Valentino2020b, Troster2021}), modified gravity (\eg \citealt{Ilic2019}), and others (\eg \citealt{Nunes2021}) However, none of the proposed solutions have gained wide acceptance so far. 
One possible solution is the decaying dark matter (DDM) model (\eg \citealt{Doroshkevich1989, Aoyama2014, Audren2014, Enqvist2015, Poulin2016, Enqvist2020, Xiao2020, Guillermo2021, Chen2021, Hubert2021, Brinch2022, Fuss2022, Guillermo2022, Mau2022, McCarthy2022, Simon2022, Bucko2023}). 
The DDM model has two features that reduce the cosmic structure growth and could naturally explain the $S_8$ tension. First, the cosmic structure growth is reduced by the decay of massive dark matter (DM) into lighter or massless daughter particles. Second, the DDM model can only change the structure formation scenario at low redshift, which is probed by clusters of galaxies, lensing, and so on, without modifying the scenario at high redshift ---which can be probed by CMB--- before dark matter decays. 

While several scenarios can be considered for the decay modes, 
we consider two scenarios: one scenario in which DM decays into a form of noninteracting dark radiation (DR), hereafter referred to as the $\Lambda$DDM1 model (\eg \citealt{Audren2014, Enqvist2015, Poulin2016, Enqvist2020}) and another scenario, where DM decays into one massless DR component and one massive warm DM (WDM) particle (hereafter $\Lambda$DDM2 model) that interacts only through gravity with the standard model particles (\eg \citealt{Aoyama2014, Guillermo2021, Guillermo2022}).
The $\Lambda$DDM1 model can be parameterized with the decay rate, $\Gamma$ ($\Gamma^{-1}$ represents the  lifetime of DM), in addition to the cosmological parameters under the standard $\Lambda$CDM model. The DDM2 model can be parameterized with two parameters: $\Gamma,$ and $\varepsilon,$ which represents the mass-energy fraction transferred to the massless component of DR \citep{Aoyama2014, Guillermo2021, Guillermo2022}. 
Many studies have been performed for the simplest scenario, that is, the $\Lambda$DDM1 model, and tight constraints have been put on the 
lifetime of the DM particle of $\Gamma^{-1}$ > 160 Gyr (\eg \citealt{Audren2014, Enqvist2015, Poulin2016, Enqvist2020}), which is much greater than the current age of the Universe. 
Fewer studies have considered the extended DDM models, such as the $\Lambda$DDM2 scenario. For example, \cite{Guillermo2022} showed that the $\Lambda$DDM2 can fully explain the low-$S_8$ measurement with a shorter lifetime of $\Gamma^{-1} \simeq 55 \, (\varepsilon / 0.007)^{1.4}$ Gyr using the \planck\ CMB measurement \citep{Planck2020VI} combined with the baryonic acoustic oscillations (BAO) measurements from the 6-degree Field Galaxy Survey (6dF) \citep{Beutler2011}, the Sloan Digital Sky Survey (SDSS) \citep{Ross2015}, the Baryon Oscillation Spectroscopic Survey (BOSS) \citep{Alam2017}, and the Extended Baryon Oscillation Spectroscopic Survey (eBOSS) \citep{Blomqvist2019, Sainte2019}, including the SNIa measurement from Pantheon \citep{Scolnic2018}. A further study was performed by \cite{Guillermo2021}, who included KiDS and DES weak-lensing data \citep{Abbott2018, Joudaki2020, Heymans2021}, with the authors concluding that there is a preference for the $\Lambda$DDM2 model over the standard $\Lambda$CDM model when KiDS and DES weak-lensing data are included, without degrading the fit to other cosmological datasets such as CMB, BAO, and SNIa measurements. 

In this paper, we use the thermal Sunyaev-Zel'dovich (SZ) effect \citep{Sunyaev1970, Sunyaev1972} as a probe for the $S_8$ tension. To investigate the $S_8$ parameter, the SZ effect, which is caused by the inverse Compton scattering of CMB photons by hot electrons along the line of sight, is a useful probe because the SZ power spectrum, $C_{\ell}^{\rm SZ}$, is sensitive to the $\sigma_8$ and $\Omega_{\rm m}$ cosmological parameters as $C_{\ell}^{\rm SZ} \propto \sigma_8^8 \Omega_{\rm m}^3$ \citep{Planck2016XXII, Salvati2018}. The $S_8$ tension was first revealed by cosmological analysis of the SZ cluster number counts from \planck\ \citep{Planck2014XX}. Recent SZ results \citep{Planck2016XXII, Tanimura2022} showed a similar $S_8$ value of $S_8 = 0.764 \, _{-0.018}^{+0.015}$ to the weak-lensing and cluster-count observations, confirming a tension with the \planck\  CMB result. 
As a solution for the $S_8$ tension, we assume that the origin of the tension is caused by the DDM and constrain the cosmological parameters and the DDM models using the latest SZ power spectrum measurement in \cite{Tanimura2022} (hereafter T22). 
In our cosmological analyses, this SZ measurement is combined with other low-redshift probes that constrain the expansion history of the Universe, such as the BAO measurements from 6dFGS at $z$ = 0.106 \citep{Beutler2011}, SDSS DR7 at $z$ = 0.15 \citep{Ross2015}, and BOSS DR12 at $z$ = 0.38 0.51, 0.61 \cite{Alam2017} as well as the Pantheon SNIa catalog \citep{Scolnic2018} (BAO+SNIa).
The present paper is structured as follows. Section \ref{sec:theory} describes the model of the SZ power spectrum, including the DDM model. Section \ref{sec:analysis} presents the constraint we place on the DDM model as a result of our cosmological analysis. We finally end with the conclusions of this study in Sect. \ref{sec:summary}.


\section{\textbf{Theoretical background}}
\label{sec:theory}

In this section, we first describe the theoretical prescription of the SZ effect, in particular the SZ angular power spectrum under the $\Lambda$CDM model by following T22, and then modify it to include the DDM models. 

\subsection{Compton $y$ parameter}

The Compton $y$ parameter is proportional to the line-of-sight integral of electron pressure, $P_{\rm e}=n_{\rm e}k_{\rm B}T_{\rm e}$, where $n_{\rm e}$ is the physical electron number density,  $k_{\rm B}$ is the Boltzmann constant, and $T_{\rm e}$ is the electron temperature. In an angular direction, $\bm{\hat{n}}$, this Compton parameter is expressed as 
\beq
    y(\bm{\hat{n}}) = \frac{\sigma_{\rm T} }{m_{\rm e} {\it c}^2} 
    \int P_{\rm e}(\bm{\hat{n}}) \, \der l\ ,
\label{eq1}
\eeq
where $\sigma_{\rm T}$ is the Thomson cross section, $m_{\rm e}$ is the mass of one electron, $c$ is the speed of light, and $l$ is the {physical} distance. 
The change to the CMB temperature by the SZ effect, $\Delta T$, at frequency $\nu$ is given by 
\beq
    \frac{\Delta T}{T_{\rm CMB}}(\nu,\bm{\hat{n}}) = f(x) \, y(\bm{\hat{n}}), 
\label{eq2}
\eeq
where $T_{\rm CMB}$ is the CMB temperature.
The frequency dependence of the SZ effect is included in the pre-factor $f(x)$ as 
\beq
    f(x) = x \ \mathrm{coth} \left(\frac{x}{2} \right) - 4 \quad \left(x = \frac{h \nu}{k_{\rm B} T_{\rm CMB}} \right)
\label{eq3}
\eeq
in the thermodynamic temperature unit, where $h$ is the Planck constant.

\subsection{SZ angular power spectrum}
\label{subsec:szmodel}
The SZ power spectrum can be modeled with a halo model \citep{Cooray2002}.
The halo model consists of a ``one-halo term'', which accounts for the correlation arising within an individual halo, and a ``two-halo term'', which accounts for the correlation arising due to the environment surrounding a halo \citep{Komatsu2002, Cooray2002}. 
In this paper, we consider only the one-halo term because the contribution from the two-halo term to the total SZ power spectrum  is minor \citep{Komatsu1999} at the scales we consider ($\ell > 60$). The SZ power spectrum is then given by 
\beq
    C^{SZ}_{\ell} = \int \der z \frac{\der^2 V}{\der z \der \Omega} \int \der M \frac{\der n(M,z)}{\der M} \, |\tilde{y}_{\ell}(M,z)|^2 , 
\label{eq:1h}
\eeq
where $\der^2 V / \der z \der \Omega$ is the comoving volume element per redshift per steradian, $n(M,z)$ is the comoving number density of halos of mass $M$ and redshift $z$, called the halo mass function (HMF), and $\tilde{y}_{\ell}(M,z)$ is the 2D Fourier transform of the $y$-profile of a halo, given by 
\beq
    \tilde{y}_{\ell}(M,z) = \frac{\sigma_{\rm T} }{m_{\rm e} {\it c}^2} \frac{4 \pi r_{\rm s}}{\ell^{2}_{\rm s}} \int \der x_{\rm r} x_{\rm r}^{2} \frac{\sin (\ell x_{\rm r}/\ell_{\rm s})}{\ell x_{\rm r}/\ell_{\rm s}} P_{\rm e}(x,M,z)\ ,
\label{eq:y2d}
\eeq
where $r_{\rm s}$ is the characteristic scale radius of the pressure profile and $x_{\rm r} = r/r_{\rm s}$ is the dimensionless radial scale. $d_{\rm A}$ is the angular diameter distance and $\ell_{\rm s} = d_{\rm A}/r_{\rm s}$ is the associated multipole moment. 
We integrate the contribution of halos in the redshift range from 0 to 3 and the mass range from $10^{13}$ $\msun$ to $5 \times 10^{15}$ $\msun$ as well as the scaled radial distance, $x_{\rm r}$, in the range from 0 to 5, following T22.

\subsection{Halo mass function}
We use the mass function from \cite{Tinker2008}, as in T22, in which the number of halos per unit volume is given by
\beq
    \frac{\der n}{\der M} = f(\sigma) \frac{\rho_{m,0}}{M} \frac{\der {\rm ln} \sigma^{-1}}{\der M}, 
\label{eq:nh}
\eeq
where $\rho_{m,0}$ is the matter density at $z=0$ and $f(\sigma)$ is given by 
\beq
    f(\sigma) = A \left[ 1 + \left( \frac{\sigma}{b} \right)^{-a} \right] {\rm exp} \left( - \frac{c}{\sigma^2} \right),
\label{eq:fsigma}
\eeq
where $A, a, b, c$ are constants calibrated with simulations in \cite{Tinker2008}. Here, $\sigma$ is the standard deviation of density perturbations in a sphere of radius $R = (3 M / 4 \pi \rho_{m,0} )^{1/3}$, and is given by
\beq
    \sigma^2 = \frac{1}{2 \pi^2} \int \der k k^2 P_{\rm m} (k,z) |W(kR)|^2, 
\label{eq:sigma}
\eeq
where $W(kR)$ is the window function of a spherical top hat of radius $R$.

\subsection{Universal pressure profile}
For the electron pressure profile, we use the model from \cite{Planck2013V}, which is used as a fiducial model in T22, and adopts the ``universal'' pressure profile \citep[UPP;][]{Nagai2007}, which is a form of the generalized \citeauthor*{Navarro1997} (NFW; \citeyear{Navarro1997}) profile (gNFW), 
\beq
    \mathbb{P} (x_{\rm 500}) = \frac{P_{0}}
        {(c_{500} x_{\rm 500})^{\gamma} [1 + (c_{500} x_{\rm 500})^{\alpha}]^{(\beta-\alpha)/\gamma} }\ .
\label{eq:gnfw}
\eeq
Here, $x_{\rm 500} = r/R_{500}$ and we remind the reader that $R_{500}$ denotes $500$ times the critical density. The model is defined by the following parameters: $P_{0}$, a normalization; $c_{500}$, a concentration parameter defined at a characteristic radius $R_{500}$; and the slopes in the central $(x_{\rm 500} \ll 1/c_{500})$, intermediate $(x_{\rm 500} \sim 1/c_{500}),$ and outer regions $(x_{\rm 500} \gg 1/c_{500})$, which are given by $\gamma$, $\alpha,$ and $\beta$, respectively. The scaled pressure profile for a halo with $M_{500}$ and $z$ is
\beq
    \frac{P(r)}{P_{500}} = \mathbb{P}(x_{\rm 500}), 
\label{eq:px}
\eeq
with 
\beq
\begin{aligned}
    P_{500} & = 1.65 \times 10^{-3} \left[\frac{H(z)}{H_0}\right]^{8/3} \\
        & \times \left[ \frac{ (1-b) \, M_{500}}{3 \times 10^{14} \, (h/0.7)^{-1} \msun} \right]^{2/3+ \alpha_{\rm p}} 
        \left(\frac{h}{0.7}\right)^{2} \rm \ keV \ cm^{-3}, 
\end{aligned}
\label{eq:p500}
\eeq
where $H(z)$ is the Hubble parameter at redshift $z$ and $H_0=100h\,\mathrm{km s^{-1} Mpc^{-1}}$ is the present value.  $P_{500}$ is the characteristic pressure reflecting the mass variation expected in a self-similar model of pressure evolution when $\alpha_{\rm p} = 0$, purely based on gravitation \citep{Arnaud2010}. 
Deviation from the self-similarity appears in a variation of the scaled pressure profile, given by 
\beq
    \frac{P(r)}{P_{500}} = 
        \mathbb{P}(x) \left[ \frac{ (1-b) \, M_{500}}{3 \times 10^{14} \, (h/0.7)^{-1} \msun} \right]^{\alpha_{\rm p}}, 
\label{eq:pr}
\eeq
expressed as a function of $M_{500}$. Here, $b$ is the hydrostatic mass bias. We note that we use the pressure model in \cite{Planck2013V} in which $M_{500}$ alternatively corresponds to the hydrostatic mass, $(1-b) \, M_{500}$. For the mass bias, we adopt $ (1-b) = 0.780 \pm 0.092$ derived from the Canadian Cluster Comparison Project  (\citealt{Hoekstra2015}; CCCP), which is consistent with most of the results from hydrodynamic simulations \citep{Gianfagna2021}. For the parameters of the generalized NFW electron pressure profile, we adopt the best-fit values of $[P_{0},c_{500}, \gamma, \alpha, \beta] = [6.41,1.81,0.31,1.33,4.13]$, estimated from 62 massive nearby clusters ($10^{14.4} < M_{500} < 10^{15.3} \msun$) using the \planck\ SZ and \xmm\ X-ray data in \cite{Planck2013V}. We also adopt $\alpha_{\rm p} = 0.12$ from the measurement in \cite{Arnaud2010}. 

\subsection{Decaying dark matter model implementation}
\label{subsec:ddmmodel}

We include the DDM model based on the public code provided by \cite{Guillermo2021}\footnote{https://github.com/PoulinV/class\_majoron} (hereafter G21), in which the DDM model is implemented in the modified version of the Boltzmann code, CLASS \citep{Blas2011, Lesgourgues2011}. 
In this code, the 1-body ($\Lambda$DDM1 model) and 2-body ($\Lambda$DDM2 model) decaying DM models are implemented by including two additional parameters with respect to $\Lambda$CDM; one is the decay rate, $\Gamma$, and the other is the mass-energy fraction transferred to DR, $\varepsilon = (1/2)(1 - m_{\rm WDM}^2 / m_{\rm DDM}^2)$, where $0 \leq \varepsilon \leq 1/2$. In the two extreme cases, $\varepsilon$ = 0 corresponds to the standard $\Lambda$CDM case with no DM decay and $\varepsilon$ = 1/2 to the $\Lambda$DDM1 case. The values in between correspond to the $\Lambda$DDM2 case.
\cite{Guillermo2021} introduced a new approximation scheme that allows one to accurately and quickly compute the dynamics of the WDM linear perturbations by treating the WDM species as a viscous fluid and included the  background evolution of WDM and, in particular, density perturbations, which were not included in previous studies \citep{Vattis2019, Haridasu2020, Clark2021}. In our analyses, we use the DDM model including this new scheme. 

In the $\Lambda$DDM1 model (DM $\rightarrow$ DR), the DM decay reduces the matter content in the Universe, causing a reduction in the expansion rate compared to the $\Lambda$CDM model, as shown in Fig. 1 of G21. This reduces the growth of matter-density fluctuations at small scales compared to the $\Lambda$CDM model, as shown in Fig. 4 of G21. The amount of matter power suppression increases as the DM lifetime becomes shorter.

In the $\Lambda$DDM2 model (DM $\rightarrow$ WDM+DR), the WDM component partially contributes to the matter density, leaving the expansion rate almost unchanged compared to the $\Lambda$CDM model, as shown in Fig. 1 of G21. However, as in the $\Lambda$DDM1 model,  this component suppresses the growth of matter-density fluctuations due to the free-streaming of WDM at small scales compared to the $\Lambda$CDM case, as shown in Fig. 4 of G21. Also similarly to the $\Lambda$DDM1 model, the amount of matter power suppression increases as the DM lifetime becomes shorter, and the scale of the power suppression is determined by the free-streaming length of WDM, similarly to that induced by massive neutrinos.

In order to include these DDM models in the model of SZ power spectrum, we modified the comoving volume, HMF, and the pressure profile of the halo in Eq.\,\ref{eq:1h}.
First, the comoving volume at a given redshift depends on the background evolution of assumed cosmological models, and the model including the DDM can be directly computed with G21. 

Second, we consider the modification of the HMF. 
The theoretical prescription of the SZ power spectrum, including the DDM model, was studied in \cite{Takahashi2004}. The authors included the impact of DDM on the HMF by modifying the Press–Schechter formalism \citep{Press1974}. 
We instead use the HMF from \cite{Tinker2008} but replace the matter power spectrum in Eq.\,\ref{eq:sigma} and $\rho_{m,0}$ in Eq.\,\ref{eq:nh} with ones including the DDM model. 
The HMF in the $\Lambda$DDM1 model was tested using N-body simulations in \cite[E20]{Enqvist2020}. E20 checked the deviation between their HMF in the $\Lambda$DDM1 model and the Tinker HMF form in the \planck $\Lambda$CDM model and provided its fitting formula with an accuracy of $\sim$20\% for halos with $10^{14} - 10^{15}$ \msunh\ at $0 < z  < 1$. 
We checked our modified Tinker HMF in the $\Lambda$DDM1 model by comparing with the results of  E20. Our HMF shows a similar trend to that found by E20: the relative discrepancy between the HMFs in the $\Lambda$DDM1 and $\Lambda$CDM models increases as the halo mass increases and the rate of DDM  increases (or the DDM lifetime decreases). 
In addition, we find that our modified HMF is consistent with that of E20 to within $\sim$5\% below $10^{14}$ \msunh\ and to within $\sim$16\% in the range of $10^{14} - 10^{15}$ \msunh\ around our best-fit value of $\Gamma^{-1} \sim 220$ Gyr, which is within the accuracy of the  HMF of E20. 
With the $\Lambda$DDM2 model, a similar result using numerical simulations has not yet been published; however, our best-fit value of $\varepsilon$ is small, namely of $\sim$ 0.002, as shown later in Table~\ref{table:result} and the impact of the $\varepsilon$ parameter on the SZ power spectrum is minor, as shown in Figure.~\ref{fig:dmwdmeps}. Therefore, our modified HMF in the $\Lambda$DDM2 model would have a similar accuracy to the one in the case of the $\Lambda$DDM1 model. 

Third, we consider the modification of the pressure profile of the halo. 
Currently, the model of this pressure profile is constrained by the pressure measurements in galaxy clusters with X-ray or the SZ effect. 
For example, in \cite{Planck2013V} and \cite{Pointecouteau2021}, the SZ profiles of galaxy clusters were stacked and fitted with the gNFW model with five parameters in Eq.\,\ref{eq:gnfw}. 
We include the impact from the DDM model on the pressure profile model in \cite{Planck2013V} by modifying the time evolution in Eq.\,\ref{eq:p500} and replacing the $H(z)$ with the one including the DDM model.
The pressure profile of the halo is also given as a function of mass, and this relation might be modified when the DDM model is included. However, we assume that this is not the case. This assumption would not be valid if the DM decay were to depend its density, but the DDM model we consider here does not have such a dependency.

\begin{figure}
    \centering
    \includegraphics[width=\linewidth]{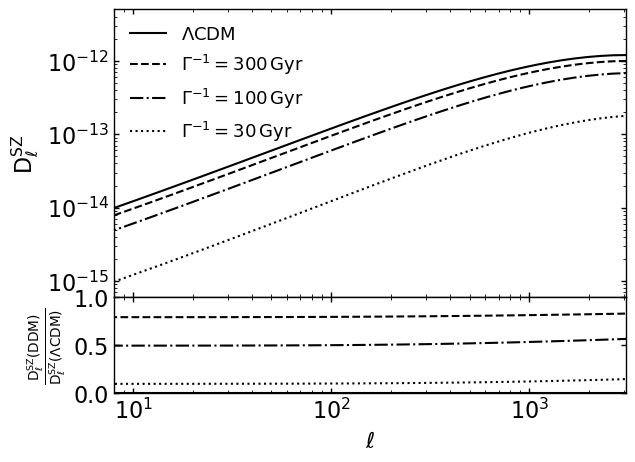}
    \caption{SZ power spectra with the $\Lambda$DDM1 model (DM $\rightarrow$ DR) with different decay lifetimes of $\Gamma^{-1}$ = 30 (dotted line), 100 (dash-dot line), and 300 (dashed line) Gyr. ``$\Lambda$CDM'' represents the $\Lambda$CDM model with stable cold dark matter (solid line). Here, the \planck\ 2018 cosmological parameters are assumed in all cases. } 
    \label{fig:dmdr}
    \end{figure}

The SZ power spectra, including these modifications, are shown in Figure.~\ref{fig:dmdr} for the $\Lambda$DDM1 model with different decay lifetimes of $\Gamma^{-1}$ = 30, 100, and 300 Gyr, which are compared with the $\Lambda$CDM model with stable CDM under the \planck\ 2018 cosmology of Table 1 in \cite{Planck2020VI}. 
As expected, the DM lifetime $\Gamma^{-1}$ determines the depth of the suppression, and the amount of power suppression increases for smaller lifetimes.

\begin{figure*}
    \centering
    \includegraphics[width=0.49\linewidth]{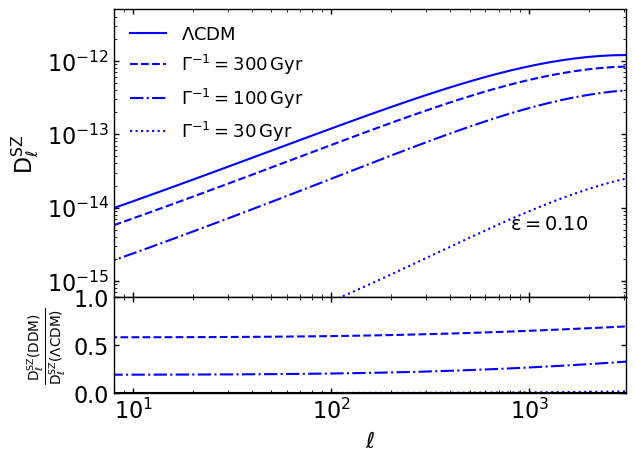}
    \includegraphics[width=0.49\linewidth]{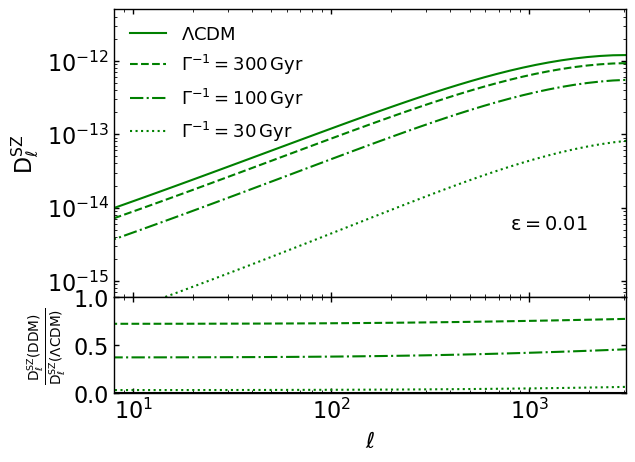}
    \caption{SZ power spectra with the $\Lambda$DDM2 model (DM $\rightarrow$ WDM+DR) with different decay lifetimes of $\Gamma^{-1}$ = 30 (dotted line), 100 (dash-dot line), and 300 (dashed line) Gyr. ``$\Lambda$CDM'' represents the $\Lambda$CDM model with stable cold dark matter (solid line). The mass-energy fractions transferred to DR are $\varepsilon$ = 0.1 (blue) in the left panel and 0.01 (green) in the right panel. Here, the \planck\
2018 cosmological parameters are assumed in all cases. }
    \label{fig:dmwdm}
    \end{figure*}

\begin{figure}
    \centering
    \includegraphics[width=\linewidth]{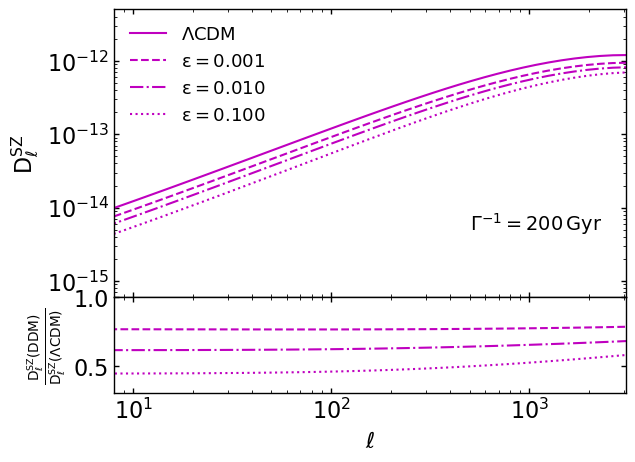}
    \caption{SZ power spectra with the $\Lambda$DDM2 model (DM $\rightarrow$ WDM+DR) for the lifetime of $\Gamma^{-1} = 200$ Gyr with different values of  $\varepsilon$ , namely, $\varepsilon$ = 0.1 (dotted line), 0.01 (dash-dot line), and 0.001 (dashed line). ``$\Lambda$CDM'' represents the $\Lambda$CDM model with stable cold dark matter (solid line). Here, the \planck\
2018 cosmological parameters are assumed in all cases.} 
    \label{fig:dmwdmeps}
    \end{figure}

We also show the SZ power spectra for the $\Lambda$DDM2 model in Figure.~\ref{fig:dmwdm} with different decay lifetimes of $\Gamma^{-1}$ = 30, 100, and 300 Gyr and also with different mass-energy fractions transferred to DR with $\varepsilon$ = 0.1 and 0.01. As in the case of the DDM1 model, the lifetime $\Gamma^{-1}$ determines the depth of the suppression, which increases for smaller lifetimes. Here, $\varepsilon$ determines the free-streaming scale of WDM and the suppression scale of the matter power spectrum, reducing the number of halos in the HMF. Because this reduction is more dramatic for massive halos, the SZ power spectrum is more suppressed at large scales, as shown in Figure.~\ref{fig:dmwdmeps}. 

\subsection{Emulator of SZ power spectrum}
\label{subsec:szemu}

We compute the $C_{\ell}^{\rm SZ}$ in our cosmological analysis with the Monte Carlo Markov chain (MCMC) in Sect. \ref{subsec:maxlkl}, and therefore the $C_{\ell}^{\rm SZ}$ computation time has to be minimized. To achieve this, we adopted the machine learning technique that uses the Random Forest algorithm developed \footnote{and distributed at https://sz-power-spectra.osups.universite-paris-saclay.fr/} by \cite{Douspis2022} and used in \cite{Gorce2022}, and modified it to build an emulator adapted to the DDM models. 
We built about 50,000 SZ power spectra with a random sampling of cosmological parameters within the ranges listed in Table.~\ref{table:cosmop} at 50 different multipoles between 2 and 3072 and used 60\%, 20\%, and 20\% of the sample for the training, validation, and test, respectively. 
We optimized the hyperparameters of the Random Forest, such as numbers of trees and depths to be 40 and 30, respectively, to maximize the final score of the prediction. This resulted in the accuracy of the constructed model being better than 1\% in our analysis range of 60 < $\ell$ < 1411, which is well below the uncertainty in the current SZ measurement of $\sim$ 30\% used in our analysis.


\section{Cosmological analysis}
\label{sec:analysis}

We use the measurement of the SZ angular power spectrum in T22 to explain the $S_8$ tension and constrain the DDM model. 
In our cosmological analyses, this SZ measurement is also combined with other low-redshift probes of BAO+SNIa.
These measurements are compared to the theoretical prescription with the DDM model described in Sect. \ref{sec:theory}.

\subsection{Compton $y$-map}
\label{subsec:data}

A new all-sky $y$ map was reconstructed in T22 from the \planck\ 2020 data release 4 \citep{Planck2020LVII}\footnote{https://pla.esac.esa.int} in HEALpix\footnote{http://healpix.sourceforge.net/} format \citep{gorski2005} with a pixel resolution of $N_{\rm side}$ = 2048 ($\sim 1.7$ arcmin). The reconstruction was performed based on the modified internal linear combination algorithm (MILCA; \citealt{hurier2013} by combining six \planck\ HFI frequency maps between 100 and 857 GHz so that the spectral response of the Compton $y$ parameter is unity. 

We use sky masks suitable for the analysis of the $y$ maps provided in the 2015 \planck\ data release \citep{Planck2016XXII}. These masks cover point sources and regions around the Galactic plane, excluding about 58\% of the sky. However, the residual extragalactic foreground emissions may remain in the $y$ map. We therefore model them including radio and infrared (IR) point sources with the \planck\ sky model \citep{Delabrouille2013} and cosmic infrared background (CIB) fluctuations from \cite{Maniyar2021}.

Our cosmological analysis uses two $y$-maps reconstructed from the first and last half-ring \planck\ frequency maps and computes their cross-power spectrum to avoid the bias induced by the noise in the auto-power spectrum. We also limit our analysis to the multipole range of 60 < $\ell$ in order to minimize the contribution from the foreground and to $\ell$ < 1411 in order to minimize the contribution from the noise.

\subsection{Maximum likelihood analysis}
\label{subsec:maxlkl}

Cosmological constraints can be obtained by fitting the SZ power spectrum measurement with the SZ and foreground models simultaneously. In our model, we consider four components: SZ, CIB, radio point sources, and IR point sources. 
We also include an instrumental noise, CN, in the model. 
Finally, the observed SZ power spectrum, $C_{\ell}^{\rm obs}$, is modeled by
\beq
    C_{\ell}^{\rm obs} = C_{\ell}^{\rm SZ} (\Theta, 1-b) + A_{\rm CIB} \, C_{\ell}^{\rm CIB} + A_{\rm IR} \, C_{\ell}^{\rm IR} + A_{\rm rad} \, C_{\ell}^{\rm rad} + A_{\rm CN} \, C_{\ell}^{\rm CN},
\label{eq:psmodel}
\eeq
where $C_{\ell}^{\rm SZ} (\Theta, 1-b)$ is the SZ power spectrum, $\Theta$ is the set of free cosmological parameters shown below in Eq.\,\ref{eq:cparams-ddm1}, $1-b$ is the hydrostatic mass bias, $C_{\ell}^{\rm CIB}$ is the CIB power spectrum, $C_{\ell}^{\rm IR}$ and $C_{\ell}^{\rm rad}$ are the IR and radio source power spectra, and $C_{\ell}^{\rm CN}$ is the empirical model for the instrumental noise. 
For the cosmological parameters, we vary
\beq
    \Theta_{\rm \Lambda DDM1} = \{\Omega_{\rm b}h^2, \ln{ (10^{10} A_{\rm s}) }, n_s, \tau_{\rm reio}, H_0, \Omega_{\rm dcdm}^{\rm ini}h^2, \log_{10} \Gamma\} 
\label{eq:cparams-ddm1}
\eeq
in the $\Lambda$DDM1 model and 
\beq
    \Theta_{\rm \Lambda DDM2} = \{\Omega_{\rm b}h^2, \ln{ (10^{10} A_{\rm s}) }, n_s, \tau_{\rm reio}, H_0, \Omega_{\rm dcdm}^{\rm ini}h^2, \log_{10} \Gamma, \log_{10} \varepsilon\}  
\label{eq:cparams-ddm2}
\eeq
in the $\Lambda$DDM2 model, 
where $\Omega_{\rm b}h^2$ is the baryon density, $\ln{ (10^{10} A_{\rm s}) }$ is the primordial density perturbation amplitude, $n_s$ is the primordial density perturbation spectral index, $\rm \tau_{reio}$ is the Reionization optical depth, $H_0$ is the Hubble constant, $\Omega_{\rm dcdm}^{\rm ini}h^2$ is the initial matter density of decaying dark matter, $\log_{10} \Gamma$ is the decay rate in logarithmic scale, and $\log_{10} \varepsilon$ is the fractional mass energy transferred to DR in logarithmic scale. 
    
We performed the MCMC likelihood analysis using the MONTEPYTHON-v3 \citep{Audren2013, Brinckmann2019}. 
We first performed the cosmological analysis only with the high-redshift \planck\ CMB data from high-$\ell$ TT,TE,EE lite and low-$\ell$ TT, EE \citep{Planck2020VI} under the $\Lambda$DDM1 and $\Lambda$DDM2 models, respectively, without the low-redshift data of the SZ, BAO, and SNIa.
We then combined the CMB data with the low-redshift probes
and compared the results. We used flat priors for the cosmological parameters and Gaussian priors for nuisance parameters. The sampling parameters and priors used in our cosmological analysis are summarized in Table.~\ref{table:cosmop}.

\begin{table*}
\caption{Sampling parameters and priors.}
\begin{center}
\begin{tabular}{lll} \hline
Parameter & Symbol & Prior \\ \hline
Baryon density & $\Omega_{\rm b}h^2$ & [0.0212 - 0.0234] \\ 
Primordial density perturbation amplitude & $\ln{ (10^{10} A_{\rm s}) }$ & [2.94 - 3.16] \\
Primordial density perturbation spectral index & $n_s$ & [0.92 - 1.00] \\
Reionization optical depth & $\rm \tau_{reio}$ & [0.01 - 0.10] \\ 
Hubble constant & $H_0$ & [64 - 71] \\  \hline
Initial matter density of decaying dark matter & $\Omega_{\rm dcdm}^{\rm ini}h^2$ & [0.11 - 0.13] \\
Inverse of decay lifetime & $\log_{10} \Gamma$ & [-4 - 1] \\ 
Fractional mass energy transferred to DR & $\log_{10} \varepsilon$ & [-4 - $\log_{10} (0.5)$] \\  \hline
CIB contamination & $A_{\rm CIB}$ & $N$(1, 0.5) \\
IR-source contamination & $A_{\rm IR}$ & $N$(1, 0.5) \\
Radio-source contamination & $A_{\rm rad}$ & $N$(1, 0.5) \\
Noise & $A_{\rm CN}$ & $N$(1, 0.5) \\ 
Mass bias & $1-b$ & $N$(0.78, 0.092) \\ \hline
\end{tabular}
\end{center}
\footnotesize{Sampling cosmological parameters and nuisance parameters are listed. 
Nuisance parameters include $A_{\rm CIB}$, $A_{\rm IR}$, $A_{\rm rad}$, $A_{\rm CN}$, and $1-b$.
[min - max] corresponds to a flat prior with minimum and maximum values.
$N$(mean, variance) corresponds to a Gaussian prior with a given mean and variance.}
\label{table:cosmop}
\end{table*}

\subsection{Results}
\label{subsec:results}

\begin{figure}
    \centering
    \includegraphics[width=\linewidth]{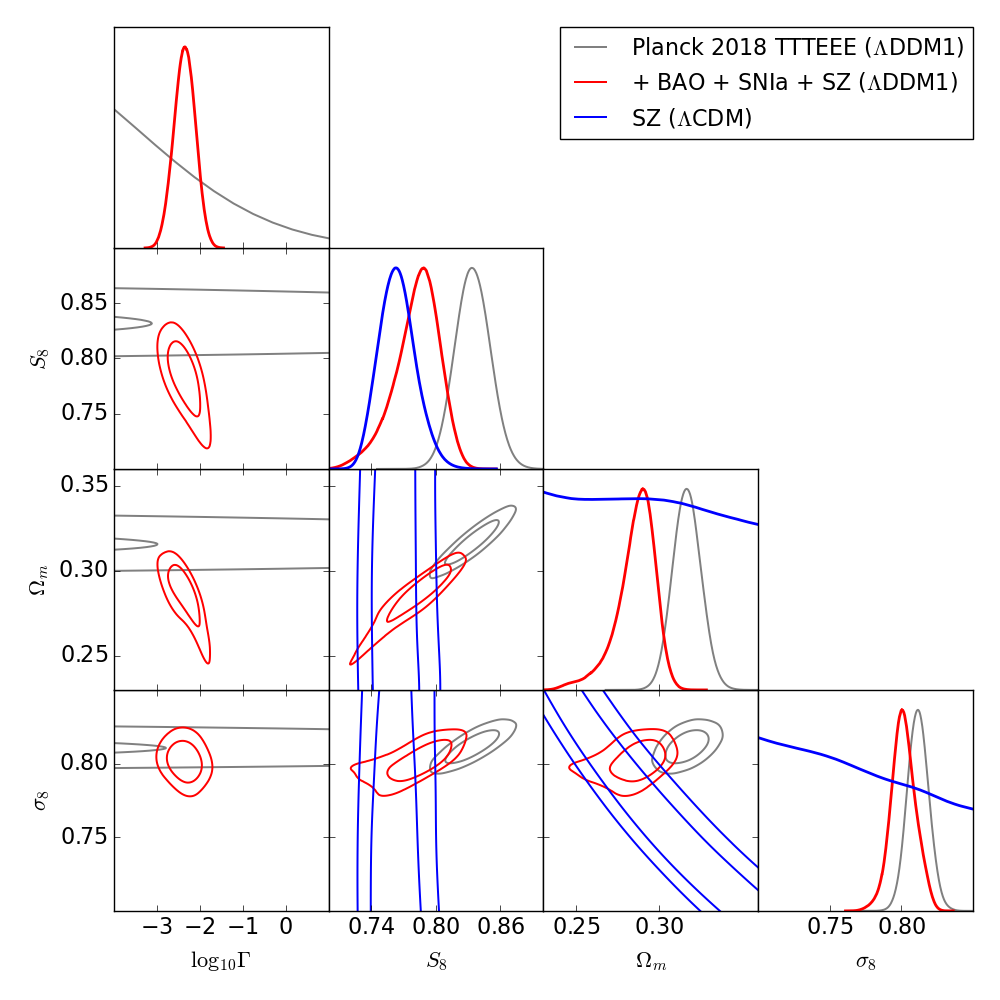}
    \caption{Posterior distributions of the cosmological parameters under the $\Lambda$DDM1 model, $\log_{10} \Gamma$, $S_8$, $\sigma_8$, and $\Omega_m$ with 68\% and 95\% confidence interval contours obtained from our cosmological analysis, when only the \planck\ CMB data are used (gray) and the \planck\ CMB data are combined with the BAO + SNIa + SZ data (red). The latest result with the SZ power spectrum from T22 is shown in blue. We note that only $\Omega_m$ and $\sigma_8$ parameters are varied in T22, while the six $\Lambda$CDM cosmological parameters are varied in our analysis as shown in Table.~\ref{table:cosmop}.}
    \label{fig:mcmc1}
    \end{figure}

Figure~\ref{fig:mcmc1} shows the result of our cosmological analysis with the $\Lambda$DDM1 model. 
In this model, DM decays into DR and $\Omega_{\rm m}$ decreases compared to that in the $\Lambda$CDM. 
Therefore, the lower $S_8$ value is mainly induced by the lower value of $\Omega_{\rm m}$, and not by that of $\sigma_8$, as shown in Figure~\ref{fig:mcmc1}. 
This figure shows the posterior distributions of the cosmological parameters, $\log_{10} \Gamma$, $S_8$, $\sigma_8$, and $\Omega_m$ with 68\% and 95\% confidence interval contours when only the \planck\ CMB data are used (gray) and the \planck\ CMB data are combined with the BAO + SNIa + SZ  (red). 
The posterior distributions of other cosmological parameters are summarized in Table.~\ref{table:result} with 68\% confidence interval, and we find that they are consistent between the two cases. 

The CMB data can constrain the DDM model through the integrated Sachs-Wolfe (ISW) effect and the lensing of the CMB as shown in Figure 5 and Figure 13 in \cite{Aoyama2014} and put a tight constraint of $\Gamma^{-1}$ > 160 Gyr on the $\Lambda$DDM1 model (\eg \citealt{Audren2014, Enqvist2015, Poulin2016, Enqvist2020}). 
In our analysis, a combination of the high-redshift CMB data and low-redshift BAO + SNIa + SZ data provides the best-fit value on the DM decay lifetime of  
\beq
    \Gamma^{-1} = 220.5_{-91.5}^{+155.8} \; {\rm [Gyr]} \quad ({\rm \Lambda DDM1}) , 
\label{eq:tau1}
\eeq
showing that the value of $\Gamma^{-1} \sim 220$ Gyr is better able to resolve the $S_8$ tension between the \planck\ CMB measurement and the SZ measurement. 
Given the uncertainty on our measurement, we also constrained a lower bound on the DM decay lifetime of $\sim$ 38 Gyr at 95\% confidence level but did not obtain a tighter constraint than that based on the CMB data. 
    
\begin{figure}
    \centering
    \includegraphics[width=\linewidth]{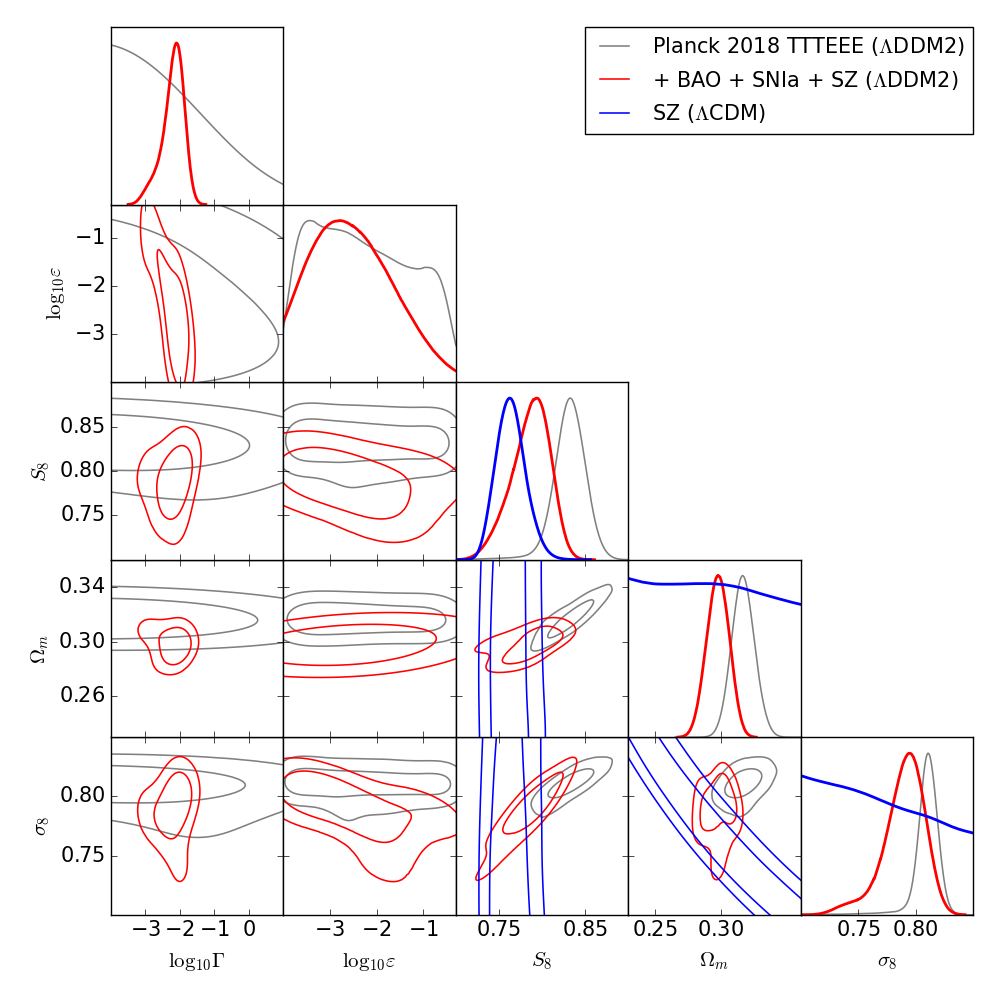}
    \caption{Posterior distributions of the cosmological parameters under the $\Lambda$DDM2 model, $\log_{10} \Gamma$, $\log_{10} \varepsilon$, $S_8$, $\sigma_8$, and $\Omega_m$ with 68\% and 95\% confidence interval contours obtained from our cosmological analysis, when only the \planck\ CMB data are used (gray) and the \planck\ CMB data are combined with the BAO + SNIa + SZ data (red). The latest result with the SZ power spectrum from T22 is shown in blue. We note that only $\Omega_m$ and $\sigma_8$ parameters are varied in T22, while the six $\Lambda$CDM cosmological parameters are varied in our analysis as shown in Table.~\ref{table:cosmop}.}
    \label{fig:mcmc2}
    \end{figure}

Figure~\ref{fig:mcmc2} shows the result of our cosmological analysis with the $\Lambda$DDM2 model.
In this model, DM decays into WDM, but $\Omega_{\rm m}$ does not decrease unlike the case in the $\Lambda$DDM1 model. (We note that the scale ranges of $\Omega_{\rm m}$ and $\sigma_8$ are the same in Figure~\ref{fig:mcmc1} and Figure~\ref{fig:mcmc2}.) Rather, the $\sigma_8$ is lowered by the matter power suppression at small scales because of the free-streaming length of WDM, which is similar to that caused by massive neutrinos. The lower $S_8$ value is therefore mainly a result of the lower value of $\sigma_8$, and not of the lower value of $\Omega_{\rm m}$, as shown in Figure~\ref{fig:mcmc2}. 
This figure shows the posterior distributions of the cosmological parameters $\log_{10} \Gamma$, $S_8$, $\log_{10} \varepsilon$, $\sigma_8$, and $\Omega_m$ with 68\% and 95\% confidence interval contours when only the \planck\ CMB data are used  (gray)\ and the \planck\ CMB data are combined with the BAO + SNIa+ SZ data (red). 
The posterior distributions of other cosmological parameters are summarized in Table.~\ref{table:result} with 68\% confidence interval, and we find that they are consistent between the two cases. 

Similarly to the case with the $\Lambda$DDM1 model, 
a combination of the high-redshift CMB data and low-redshift BAO+SNIa+SZ data provides the best-fit value on the DM decay lifetime of  
\beq
    \Gamma^{-1} = 137.3_{-56.6}^{+176.8} \; {\rm [Gyr]} \quad ({\rm \Lambda DDM2}), 
\label{eq:tau2}
\eeq
which gives a lower bound on the decay lifetime of  DM of $\sim$ 24 Gyr at 95\% confidence level. 

In summary, we find that both models give lower $S_8$ values than the value from the \planck\ CMB measurement and reconcile the $S_8$ tension. Therefore, both models may provide a solution for the tension. We also compare the best-fit $\chi^2$ values in our cosmological analyses. These are 1021.94 and 1020.67 with the $\Lambda$DDM1 and $\Lambda$DDM2 models, respectively, and we find a slight preference for the $\Lambda$DDM2 model compared to the $\Lambda$DDM1 model.

However, one of the DDM parameters, $\varepsilon$, is not well constrained in our analysis. To constrain this parameter, precise measurements of the shape of the SZ power spectrum are required as shown in Figure.~\ref{fig:dmwdmeps}. However, the uncertainty in the current SZ measurement is $\sim$ 30\%, which prevents us from determining the value of this parameter.
For this purpose, more sensitive data at small scales will be useful, such as those provided by AdvACT \cite{Henderson2016} and SPT-3G \cite{Benson2014} and the upcoming Simons Observatory \cite{Ade2019} and CMB-S4 \cite{Abazajian2019}.

\begin{table*}
\caption{Best cosmological parameter estimates under the ${\rm \Lambda DDM1}$ and ${\rm \Lambda DDM2}$ models using the \planck\ CMB data without or with SZ data.}
\begin{center}
\begin{tabular}{|l|c|c|c|c|} \hline
Parameter & \multicolumn{2}{|c|}{$\Lambda$DDM1} & \multicolumn{2}{|c|}{$\Lambda$DDM2}\\ \hline
 & \planck\ CMB & +BAO+SNIa+SZ & \planck\ CMB & +BAO+SNIa+SZ \\ \hline
100 $\Omega_{\rm b}h^2$ & 0.0223 $\pm$ 0.0002 & 0.0224 $\pm$ 0.0001 & 0.0223 $\pm$ 0.0002 & 0.0226 $\pm$ 0.0002 \\ 
$\ln{ (10^{10} A_{\rm s}) }$ & $3.045_{-0.016}^{+0.017}$ & $3.045_{-0.018}^{+0.021}$ & $3.045_{-0.015}^{+0.016}$ & $3.042_{-0.016}^{+0.020}$ \\
$n_s$ & 0.9642 $\pm$ 0.0045 & $0.9683_{-0.0041}^{+0.0053}$ & $0.9641_{-0.0043}^{+0.0044}$ & $0.9726_{-0.0069}^{+0.0041}$ \\ 
$\rm \tau_{reio}$ & $0.0542_{-0.0078}^{+0.0081}$ & $0.0566_{-0.0084}^{+0.0099}$ & $0.0560_{-0.0083}^{+0.0096}$ & $0.0560_{-0.0085}^{+0.0082}$ \\ 
$H_0$ &  67.2 $\pm$ 0.6 & $68.3_{-0.5}^{+0.6}$ & 67.2 $\pm$ 0.6 & $68.6_{-0.6}^{+0.7}$ \\ 
$\Omega_{\rm dcdm}^{\rm ini}h^2$ & 0.1202 $\pm$ 0.0014 & 0.1187 $\pm$ 0.0012 & 0.1202 $\pm$ 0.0014 & $0.1172_{-0.0014}^{+0.0016}$ \\
$\log_{10} \Gamma$ & Not preferred & -2.34 $\pm$ 0.23 & Not constrained & $-2.14_{-0.36}^{+0.23}$ \\ 
$\log_{10} \varepsilon$ & - &  - & Not constrained & $-2.68_{-0.87}^{+1.07}$ \\  
$\Omega_{\rm m}$ & $0.3168_{-0.0087}^{+0.0089}$ & $0.288_{-0.012}^{+0.009}$ & $0.3167_{-0.0084}^{+0.0087}$ & 0.298 $\pm$ 0.009 \\  
$\sigma_8$ & $0.8116_{-0.0076}^{+0.0078}$ & $0.8013_{-0.0082}^{+0.0091}$ & $0.8108_{-0.0085}^{+0.0080}$ & $0.7932_{-0.0198}^{+0.0139}$ \\
$S_8$ & 0.834 $\pm$ 0.017 & $0.7852_{-0.023}^{+0.016}$ & 0.833 $\pm$ 0.017 & $0.7901_{-0.024}^{+0.019}$ \\  \hline
\end{tabular}
\end{center}
\label{table:result}
\end{table*}

\subsection{Systematic effects}
\label{subsec:systematics}

Finally, we consider systematic uncertainties in our analysis. As discussed in \cite{Planck2016XXII} and T22, 
the mass bias and the pressure profile model are not well constrained, which affects the cosmological analysis with the SZ power spectrum. We summarize the results in Table~\ref{table:ddm1bias} and Table~\ref{table:ddm2bias} for the $\Lambda$DDM1 and $\Lambda$DDM2 models.

First, in order to investigate the impact on our DDM constraints from the mass bias, we replaced the mass bias prior on the  CCCP with two others: one from the ``weighting the giants'' weak lensing measurements (\citealt{Linden2014}; WtG) and the other from cosmological hydrodynamical simulations (\citealt{Biffi2016}; BIFFI). 
Table~\ref{table:ddm1bias} shows a comparison of the effects of these two mass bias priors on $\Gamma_{-1}$ and its lower bound at 95\% confidence level. The result shows that the $\Gamma^{-1}$ value increases as the mass bias increases (or $1-b$ decreases). 
The reason is that the amplitude of the SZ power spectrum model shifts downwards when the mass bias increases, as in Eq.\,\ref{eq:p500}, which reduces the $S_8$ discrepancy between the \planck\ CMB and SZ results and thus requires a smaller amount of DDM or, in other words, a longer decay lifetime. 

In addition, we investigated the impact of the pressure profile model  on our DDM constraints by replacing P13 with two other models: one based on the combination of \xmm\ measurements and numerical simulations (\citealt{Arnaud2010}; A10) and the other based on the analysis with combined data from \planck\ and ACT (\citealt{Pointecouteau2021}; PACT21). 
Table~\ref{table:ddm1bias} shows a comparison of the effects of these models on $\Gamma^{-1}$ and its lower band at 95\% confidence level. The result shows that the $\Gamma^{-1}$ value of PACT21 is higher than the others.  The reason is that the amplitude of the PACT21 pressure profile is slightly lower than the others, as shown in Fig. 5 of \cite{Pointecouteau2021}. This shifts the amplitude of the SZ power spectrum model downwards, reduces the $S_8$ discrepancy between the \planck\ CMB and SZ results, and requires less DDM. 

\begin{table}
\caption{$\Gamma^{-1}$ constraints obtained with different mass bias priors and pressure profile models under ${\rm \Lambda DDM1}$.} 
\label{table:ddm1bias}      
\centering                          
\begin{tabular}{c c c}        
\hline\hline                 
Mass bias prior & $1-b$ & $\Gamma^{-1} \rm [Gyr]$ (lower bound) \\  
\hline                        
CCCP (fiducial) & $0.780 \pm 0.092$ & $220.5_{-91.5}^{+155.8}$ (> 37.5) \\ 
WtG & $0.688 \pm 0.072$ & $255.1_{-108.8}^{+119.8}$ (> 37.5) \\  
BIFFI & $0.877 \pm 0.015$ & $156.1_{-51.2}^{+82.2}$ (> 53.7) \\  
\hline\hline                               
Pressure model & & $\Gamma^{-1} \rm [Gyr]$ (lower bound) \\  
\hline                        
P13 (fiducial) & & $220.5_{-91.5}^{+155.8}$ (> 37.5) \\ 
PACT21 & & $276.1_{-131.5}^{+143.4}$ (> 13.1) \\  
A10 & & $209.3_{-73.3}^{+110.8}$ (> 62.7) \\  
\hline 
\hspace{0.5cm}
\end{tabular}
\caption{$\Gamma^{-1}$ constraints obtained with different mass bias priors and pressure profile models under ${\rm \Lambda DDM2}$.} 
\label{table:ddm2bias}      
\centering                          
\begin{tabular}{c c c}        
\hline\hline                 
Mass bias prior & $1-b$ & $\Gamma^{-1} \rm [Gyr]$ (lower bound) \\  
\hline                        
CCCP (fiducial) & $0.780 \pm 0.092$ & $137.3_{-56.6}^{+176.8}$ (> 24.1) \\  
WtG & $0.688 \pm 0.072$ & $190.1_{-81.3}^{+157.8}$ (> 27.5) \\  
BIFFI & $0.877 \pm 0.015$ & $102.2_{-37.4}^{+133.6}$ (> 27.4) \\  
\hline\hline                               
Pressure model & & $\Gamma^{-1} \rm [Gyr]$ (lower bound) \\  
\hline                        
P13 (fiducial) & & $137.3_{-56.6}^{+176.8}$ (> 24.1) \\  
PACT21 & & $181.9_{-87.8}^{+198.9}$ (> 6.3) \\  
A10 & & $141.1_{-54.9}^{+164.6}$ (> 31.3) \\  
\hline 
\end{tabular}
\end{table}

\section{Summary and conclusion}
\label{sec:summary}

To solve the $S_8$ tension between the \planck\ CMB measurement and low-redshift probes, we extend the $\Lambda$CDM model, including a decaying dark matter (DDM) model. Two DDM models are tested in our study: one DDM model ($\Lambda$DDM1) where DM decays into a form of noninteracting dark radiation (DR), which is parameterized with the decay rate, $\Gamma$, and another model ($\Lambda$DDM2), where the DM decays into warm dark matter (WDM) and DR, which is parameterized with the decay rate, $\Gamma$, and the mass-energy fraction transferred to the massless component, $\varepsilon$. 
For the low-redshift probe, we use the Sunyaev Zel'dovich effect and compute the impact of  DDM on the SZ power spectrum by varying the DDM parameters, including the background evolution in cosmology and nonlinear prescription in the halo mass function. The result shows the suppression of the SZ power spectrum relative to the $\Lambda$CDM model due to DDM. As one expects, the suppression becomes more remarkable as the decay time becomes shorter. We combine this SZ data with other low-redshift probes, namely BAO+SNIa, which constrain the expansion history of the Universe.

We performed a cosmological analysis and compared the results when only the \planck\ CMB data are used and when the \planck\ CMB data are combined with the BAO, SNIa, and SZ data from T22. Under the $\Lambda$DDM1 model, the result shows a preference for $\Gamma^{-1} = 220.5_{-91.5}^{+155.8}$ Gyr to resolve the tension between the \planck\ CMB and SZ measurements on the $S_8$ parameter, in agreement with literature supporting $\Gamma^{-1}$ > 160 Gyr. Given the uncertainty on our measurement, we also set a lower bound on the DM decay lifetime of $\sim$ 38 Gyr at 95\% confidence level. 
We also performed a cosmological analysis under the $\Lambda$DDM2 model using the \planck\ CMB and BAO+SNIa+SZ measurements. The result shows a preference for $\Gamma^{-1} = 137.3_{-56.6}^{+176.8}$ Gyr to resolve the $S_8$ tension. This result provides a lower bound on the DM decay lifetime of $\sim$ 24 Gyr at 95\% confidence level. 
As a result, both models give lower $S_8$ values than the value from the \planck\ CMB measurement and reconcile the $S_8$ tension. Thus, both models may provide a solution for the tension. In addition, we compared the best-fit $\chi^2$ values in these cosmological analyses and find a slight preference for the $\Lambda$DDM2 model compared to the $\Lambda$DDM1 model.

We also checked the systematic uncertainty in our analysis with the SZ power spectrum, which includes the impact of the mass bias and pressure profile model. The lower bound changes depending on the choice of these models. This indicates that the DDM model can only be more accurately constrained if the systematic uncertainties originating from the internal structures in galaxy clusters ---such as contributions of nonthermal pressure and baryonic feedback effects--- are well understood and modeled. 

Finally, current studies based on weak lensing and galaxy clustering as low-redshift probes indicate that a lower $S_8$ value may be caused by a lower $\sigma_8$, and not a lower $\Omega_m$ \citep{snowmass2021-III}. To investigate this using the SZ signal, additional high-resolution and high-sensitivity data will be needed, such as those of AdvACT and SPT-3G and the upcoming Simons Observatory and CMB-S4.


    
\begin{acknowledgements}
This research has been supported by the funding for the Baryon Picture of the Cosmos (ByoPiC) project from the European Research Council (ERC) under the European Union's Horizon 2020 research and innovation programme grant agreement ERC-2015-AdG 695561. The authors acknowledge fruitful discussions with the members of the ByoPiC project (https://byopic.eu/team). 
This work is supported by World Premier International Research Center Initiative (WPI), MEXT, Japan. Kavli IPMU was established by World Premier International Research Center Initiative (WPI), MEXT, Japan. Kavli IPMU is supported by World Premier International Research Center Initiative (WPI), MEXT, Japan. 
\end{acknowledgements}

\bibliographystyle{aa} 
\bibliography{ddm} 

\end{document}